\documentclass[%
reprint,
superscriptaddress,
 amsmath,amssymb,
aps,
prb,
]{revtex4-1}

\usepackage[usenames,dvipsnames]{color}

\usepackage{graphicx}
\usepackage{dcolumn}
\usepackage{bm}

\newcommand{\itemph}[1]{\textit{#1}}
\newcommand{\be}{\begin{equation}}
\newcommand{\ee}{\end{equation}}
\newcommand{\bea}{\begin{eqnarray}}
\newcommand{\eea}{\end{eqnarray}}

\newcommand{\atan}{\text{atan}}

\newcommand{\ket}[1]{\lvert#1\rangle}

\newcommand{\braket}[2]{\langle#1\vert#2\rangle}

\usepackage{ulem}

\begin{document}

\title{Quasi-soliton scattering in quantum spin chains}
\date{\today}
 
\author{R. Vlijm} \email{R.P.Vlijm@uva.nl} \affiliation{Institute for Theoretical Physics, University of Amsterdam, Science Park 904, 1090 GL Amsterdam, The Netherlands}
\author{M. Ganahl} \affiliation{Institut f\"{u}r Theoretische Physik, Technische Universit\"at Graz, Petersgasse 16, 8010 Graz, Austria}
\author{D. Fioretto} \affiliation{Institute for Theoretical Physics, University of Amsterdam, Science Park 904, 1090 GL Amsterdam, The Netherlands}\affiliation{Institut f\"{u}r Theoretische Physik, Friedrich-Hund-Platz 1, 37077 G\"{o}ttingen, Germany}
\author{M. Brockmann} \affiliation{Institute for Theoretical Physics, University of Amsterdam,
  Science Park 904, 1090 GL Amsterdam, The Netherlands}
\affiliation{Max-Planck-Institut f\"{u}r Physik komplexer Systeme, N\"othnitzer Stra{\ss}e 38, 01187
  Dresden, Germany} 
\author{M. Haque} \affiliation{Max-Planck-Institut f\"{u}r Physik komplexer Systeme, N\"othnitzer Stra{\ss}e 38, 01187 Dresden, Germany}
\author{H.G. Evertz} \affiliation{Institut f\"{u}r Theoretische Physik, Technische Universit\"at Graz, Petersgasse 16, 8010 Graz, Austria}
\author{J.-S. Caux} \affiliation{Institute for Theoretical Physics, University of Amsterdam, Science Park 904, 1090 GL Amsterdam, The Netherlands}

\begin{abstract}
The quantum scattering of magnon bound states in the anisotropic Heisenberg spin chain is shown to display features similar to the scattering of solitons in classical exactly solvable models. Localized colliding Gaussian wave packets of bound magnons are constructed from string solutions of the Bethe equations and subsequently evolved in time, relying on an algebraic Bethe ansatz based framework for the computation of local expectation values in real space-time. The local magnetization profile shows the trajectories of colliding wave packets of bound magnons, which obtain a spatial displacement upon scattering. Analytic predictions on the displacements for various values of anisotropy and string lengths are derived from scattering theory and Bethe ansatz phase shifts, matching time evolution fits on the displacements. The time evolved block decimation (TEBD) algorithm allows for the study of scattering displacements from spin-block states, showing similar scattering displacement features.
\end{abstract}

\maketitle

\section{Introduction} \label{sec:intro}
The study of classical dynamics in nonlinear media has proven to be a source of astonishing
surprises over the last century. Two observations, both based on numerical simulations, have
challenged prejudices and fundamentally altered traditonal ways of thinking. First, the famous
observation of Fermi, Pasta, Ulam, and Tsingou\cite{FPU} of a simple nonlinearly-coupled set of
oscillators showing nontrivial recurrences, has shattered the long-held assumption that all nonlinear
dynamical systems ergodically explore their full phase space. Second, the pioneering numerical
analysis of Zabusky and Kruskal \cite{1965_Zabusky_PRL_15} on the Korteweg-de Vries equation
\cite{KdV} demonstrated that this equation supports excitations, which they coined `solitons',
displaying a number of surprising fundamental features. The solitons are localized in space, with a
form remaining stable under time evolution which sees them moving uniformly at a speed linearly
proportional to their amplitude. Additionally the astounding characteristic was observed of solitons
emerging intact from mutual scattering processes, during which they simply `{``pass through''} one
other without losing their identity',\cite{1965_Zabusky_PRL_15} the only effect of the collision being a relative spatial displacement as compared to their free propagation. The proper understanding of solitons in nonlinear media ultimately led to the development of the classical inverse scattering method, \cite{GardnerGreenKruskalMiura,TakhtajanBOOK} which is the overarching framework for classical integrable models.

Nonlinear classical systems find their quantum mechanical analogue in the shape of interacting many-body systems. In parallel to the classical case, some quantum models have been shown to be special, in the sense of being exactly solvable using the quantum version of the inverse scattering method \cite{KorepinBOOK} (one might also say integrable, though the quantum notion of integrability is not as well-defined as its classical counterpart \cite{2011_Caux_JSTAT_P02023}). Fundamental representatives of this family are the Heisenberg spin chain, solved by Bethe using what is now known as the Bethe ansatz,\cite{1931_Bethe_ZP_71} along with the Lieb-Liniger model of $\delta$-interacting bosons on a line.\cite{1963_Lieb_PR_130_1} For the latter, particle-like excitations called Lieb Type I modes \cite{1963_Lieb_PR_130_2} exist due to the interparticle repulsion, along with Type II hole-like excitations visualized as holes in an effective Fermi sea. It is possible to distinguish the presence of Type I and II modes in correlated bosonic gases in optical lattices using Bragg spectroscopy.\cite{2015_Fabbri_PRA_91,arxiv_1505_08152} Quantum magnets such as the Heisenberg spin chain similarly carry particle-like magnon modes when the magnetization is close to saturation. In the limit of small magnetization, hole-like modes again appear, which in zero field are known as spinons.\cite{1981_Faddeev_PLA_85} Their dynamics can be experimentally observed using inelastic neutron scattering.\cite{2013_Mourigal_NATPHYS_9,2013_Lake_PRL_111}
The Heisenberg chain supports distinct bound states of magnons, whose dynamics
has been investigated theoretically\cite{2012_Ganahl_PRL_108} and has recently been observed experimentally.\cite{2013_Fukuhara_NATURE_502}

One could view such excitations as the quantum equivalents of classical solitons. This equivalence
is however only partial: on the one hand, these quantum mechanical modes represent exact eigenstates
and are stable under time evolution; on the other hand, being exact eigenstates of
translationally-invariant systems, they are not spatially localized. 
That said, as is usually the case in quantum mechanics, it is possible to adopt a `complementary' picture and create spatially localized excitations by forming wave packets of fundamental excitations by linearly combining states over a range of differing momenta.
Locality however comes at a price: the wave packet, mixing together states at different energies, will disperse and is thus not stable over long timescales, unlike its classical counterpart. We use the term quasisoliton for such a wave packet construction, an example of which was recently studied in the context of the Lieb-Liniger model,\cite{2012_Deguchi_PRL} while their mutual scattering has been studied in quantum spin chains.\cite{GanahlHaqueEvertz_arxiv13, 2012_WoellertHonecker_PRB}

The spectroscopic methods traditionally employed to experimentally study condensed-matter systems typically provide momentum- and energy-resolved measurements. 
However, current experimental developments provide motivation to obtain a better theoretical understanding of spatially localized dynamics. Time-resolved experiments are now able to track quantum many-body systems at timescales smaller than relaxation timescales, particularly in experiments with ultracold atoms,\cite{Cold-atom_reviews_expts, 2013_Fukuhara_NATURE_502, 2013_Fukuhara_NaturePhysics} but also in NMR setups,\cite{nmr} and potentially in pump-probe spectroscopy experiments.\cite{expts_pump-probe}  As a result, 
 non-dissipative dynamics of many-body systems out of equilibrium is now a rapidly growing field of experimental as well as theoretical investigation.\cite{noneq_reviews} Novel cold-atom experimental techniques for spatially resolved manipulation and observation at the single-site level \cite{SingleSiteResolutionExpts,2013_Fukuhara_NATURE_502, 2013_Fukuhara_NaturePhysics} have opened the door to explicit high-resolution tracking of spatial propagation phenomena.
 Moreover, an experiment with interacting bosonic atoms has highlighted the interaction-induced
 longevity of repulsive pairs. \cite{2006_Winkler_Nature} This has motivated increased theoretical
 attention to the (anti-)bind\-ing of localized excitations and interactions of these bound
 clusters, both in itinerant systems\cite{bound_clusters_itinerant, GanahlHaqueEvertz_arxiv13} and
 in spin chains.\cite{2012_Ganahl_PRL_108, GanahlHaqueEvertz_arxiv13, 2012_WoellertHonecker_PRB,
   xxz_spinblocks} Very recently, the spatial dynamics of itinerant clusters has been studied in
 experiment. \cite{Weiss_NPhys2015} Propagation of quantum solitons in Bose-Hubbard chains has also
 been studied numerically. \cite{solitons_BoseHubbard}

In Ref.~\onlinecite{GanahlHaqueEvertz_arxiv13}, the scattering of a magnon wave packet on approximate bound eigenstates of $n$ particles was studied numerically in the Heisenberg chain.
In the present work, we present an algebraic framework and exact calculations based on Bethe ansatz. We therefore consider quantum scattering of localized excitations
over a ferromagnetic background in the Bethe ansatz solvable anisotropic spin-1/2 Heisenberg chain (XXZ model) \cite{1931_Bethe_ZP_71,1928_Heisenberg_ZP_49,1958_Orbach_PR_112}
\begin{equation}
  H = J \sum_{j=1}^{N} \left[ S^x_j S^x_{j+1}+S^y_j S^y_{j+1}+\Delta\left(S^z_j S^z_{j+1}-\frac{1}{4}\right)\right]. 
\label{eq:xxzhamiltonian}
\end{equation}
The XXZ model is experimentally realizable, for example in the setups of Refs.~\onlinecite{2013_Fukuhara_NATURE_502} and~\onlinecite{2013_Fukuhara_NaturePhysics}, which use two hyperfine states of bosonic atoms in the Mott phase to experimentally realize the spin-up and spin-down states. The effective model is a nearly isotropic ($\Delta \approx 1$) Heisenberg chain, while an experimental setup with variable anisotropy $\Delta$ is under development.\cite{Gross_Bloch_private} In addition, the XXZ model has been shown to describe Josephson junction arrays of the flux qubit type,\cite{LyakhovBruder_NJP05} and may also be realizable in optical lattices\cite{DuanDemlerLukin_PRL03} or with polaritons in coupled arrays of cavities.\cite{cca_KayAngelakis} It is conceivable that these or similar experimental setups may provide time-resolved observations of propagating and interacting localized excitations.

The parameter $J$ in Hamiltonian \eqref{eq:xxzhamiltonian} is given by the exchange interaction of
two neighboring electrons or, in the aforementioned experimental setup, by the exchange interaction
between two neighboring bosons in an optical lattice at unit filling. Its sign does not matter for
our purposes
here;\cite{GanahlHaqueEvertz_arxiv13,2012_Schneider_Nature_Physics,2010_Altman_Demler_PRA} for
definiteness, we set it to $J > 0$. We distinguish two regions for the anisotropy
parameter~$\Delta$, namely the planar $xy$ ($|\Delta| < 1$) or axial $z$ ($|\Delta| > 1$) cases. Due
to the interaction term ($\propto\Delta$) the XXZ chain displays a whole zoology of fundamental
excitations: isolated down spins can form spatially bound states (as was understood by Bethe already
in his original publication \cite{1931_Bethe_ZP_71}), whose bond size decreases as $\Delta$
increases. These excitations are often referred to as `string states', as the rapidities describing
Bethe states containing such bound multi-magnons appear as approximately equally-spaced vertical
strings in the complex plane (the precise set of available bound states depends on the value of
$\Delta$; at the isotropic point $\Delta = 1$ and in the axial regime, all string lengths are
allowed).

We construct spatially localized wave packets of $n$ bound magnons using linear combinations of these string states of length $n$ with Gaussian-distributed momenta. We call them `$n$-string wave packets'. We further demonstrate that exact methods based on the algebraic Bethe ansatz~\cite{KorepinBOOK} provide a framework to evaluate the time-dependent expectation value of the local magnetization $\langle S^z_j(t) \rangle$ algebraically, which can be used to track those localized magnon-like wave packets. We investigate their stability and mutual scattering using a combination of scattering theory, Bethe ansatz, and numerically exact calculations.

At large anisotropy $\Delta \gg 1$, magnon bound states resemble having downturned spins on neighboring sites. An $n$-string wave packet  is thus closely approximated by  a consecutive block of $n$ downturned spins. While this correspondence breaks down at smaller $\Delta$, this provides motivation to study the evolution of states with downturned spins on a consecutive block of sites. In addition, this is exactly the type of initial state prepared in experiments.\cite{2013_Fukuhara_NATURE_502, 2013_Fukuhara_NaturePhysics} 
Scattering of such blocks has been explored 
in Ref.~\onlinecite{GanahlHaqueEvertz_arxiv13}, where a spatial displacement of two sites was observed for the block at several $\Delta$, and explained at large $\Delta$ in terms of energy conservation.  In the present work, we connect scattering phase shifts with trajectory displacements in order to provide a Bethe ansatz derivation of the observed displacements.

The paper is organized as follows. In Sec.~\ref{sec:xxz} we will introduce the concepts of string solutions and the scattering phase shifts associated with $n$-strings, along with details on the algebraic Bethe ansatz\cite{KorepinBOOK} based evaluation of the time-dependent expectation value of the local magnetization $\langle S^z_j(t) \rangle$, which can be used to track localized magnon-like excitations of the spin chain. 
In Sec.~\ref{sec:wave packets}, we elaborate on the construction, stability and time evolution of scattering of $n$-string wave packets. We also compare with the stability of consecutive-site spin blocks.
In Sec.~\ref{sec:displacement} we consider the scattering trajectory displacements, derive analytical results for them and compare with numerical measurements. The $\Delta\to\infty$ limit is treated analytically, comparing with numerical data for spin block initial states.
The appendices provide details involving scattering theory and details on obtaining the phase shift directly from the phase of the time-evolving wave function.

\section{Time evolution from Bethe ansatz}\label{sec:xxz}
In this section we present the basic formulas of the Bethe ansatz for the spin-1/2 XXZ model and explain the concept of string solutions and the scattering phase shift associated with $n$-strings. In Subsec.~\ref{sec:timeevoABA}, expressions are given for the time evolved expectation value of the local magnetization by using results from algebraic Bethe ansatz. The method used here relies on the availability of determinant expressions for matrix elements between Bethe states.\cite{1989_Slavnov_TMP_79,1990_Slavnov_TMP_82,1999_Kitanine_NPB_554} This last subsection is relatively technical and could be skipped on first reading.

\subsection{Coordinate Bethe ansatz for the XXZ model}
The eigenstates of the XXZ spin chain~\eqref{eq:xxzhamiltonian} can be constructed via the Bethe ansatz,\cite{1931_Bethe_ZP_71,1958_Orbach_PR_112} and have the form
\begin{equation}
	|\{\lambda\}\rangle = \!\!\!\! \sum_{{j_1 < \ldots < j_M}}\!\! \sum_{Q}A_Q(\{\lambda\})\prod_{a=1}^M e^{i j_a p(\lambda_{Q_a})} S_{j_a}^-\left|\uparrow\uparrow\ldots\uparrow\right\rangle ,
	\label{eq:bethestates}
\end{equation}
where $M$ denotes the number of downturned spins and therefore fixes the magnetization. The sum over $Q$ is a sum over all permutations of $M$ objects and the amplitudes $A_Q$ are related to the scattering phases. The set of $M$ complex rapidities $\{\lambda\}\equiv \{\lambda_j\}_{j=1}^M$ completely determines the Bethe state and is simply related to the physical energy and momentum. By imposing periodic boundary conditions, i.e.~$S_{N+1}^\alpha = S_1^{\alpha}$ for $\alpha=x,y,z$ in the Hamiltonian \eqref{eq:xxzhamiltonian}, each set of rapidities $\{\lambda\}$ corresponding to an $M$-magnon eigenstate must obey Bethe equations,
\begin{equation}
  \left( \frac{\phi_{1}(\lambda_j)}{\phi_{-1}(\lambda_j)} \right)^N = \prod_{\substack{k=1 \\ k\neq j}}^M \frac{\phi_2 (\lambda_j-\lambda_k)}{\phi_{-2} (\lambda_j-\lambda_k)} \,.
\label{eq:betheeqprod}
\end{equation}
The different definitions of $\phi_n(\lambda)$, $\theta_n(\lambda)$ and $\zeta$ for various regions in anisotropy $\Delta$ are given in Tab.~\ref{table:kerneldef}. Within Bethe ansatz, the momenta of single downturned spins can be parametrized in terms of rapidities,
\begin{equation}
  p(\lambda) = -i \ln \left[ \frac{\phi_1(\lambda)}{\phi_{-1}(\lambda)} \right] = \pi - \theta_1(\lambda)\,.
\end{equation}

\begin{table}
\begin{ruledtabular}
\begin{tabular}{l|ccc}
&$\zeta$ & $\phi_n(\lambda)$ & $\theta_n(\lambda)$\\
\hline
$ |\Delta| < 1$ & $\text{acos}\,(\Delta)$ & $\text{sinh}\left(\lambda+\frac{i n \zeta}{2}\right)$ & $2\,\atan\left(\cfrac{\tanh(\lambda)}{\tan(\frac{n \zeta}{2})}\right) $\\
$\phantom{||} \Delta=1$ & $-$ & $\lambda+\frac{i n}{2}$ & $2\,\atan\left(\cfrac{2\lambda}{n}\right)$\\
$\phantom{||} \Delta > 1$ & $\text{acosh}\,(\Delta)$ & $\text{sin}\left(\lambda+\frac{i n \zeta}{2}\right)$ & $2\,\atan\left(\cfrac{\tan (\lambda)}{\tanh(\frac{n \zeta}{2})}\right)$
\end{tabular}
\end{ruledtabular}
\caption{Definitions of functions appearing in Bethe ansatz for different regions in anisotropy $\Delta$.}
\label{table:kerneldef}
\end{table}

By invoking Schr\"odinger's equation $H|\{\lambda\}\rangle=E|\{\lambda\}\rangle$ for a Bethe state consisting of a single downturned spin, the magnon dispersion relation
\begin{equation}\label{eq:disp_rel}
  E(p)=J(\cos(p)-\Delta)
\end{equation}
is easily derived. In the case of just two single magnons, their scattering phase shift $\chi$ can be obtained from the permutation of two magnons in the Schr\"odinger equation,
\begin{equation}
  \frac{A_{Q^\prime}}{A_{Q}} = -\frac{1+e^{i(p_1+p_2)}-2\Delta e^{i p_2}}{1+e^{i(p_1+p_2)}-2\Delta e^{i p_1}} = -e^{i \chi}\,,
\label{eq:BAphaseAoverA}
\end{equation}
where $Q$ is the identity and $Q'$ interchanges the two indices $1$ and $2$. Furthermore, the magnon momenta $p_1 = p(\lambda_1)$ and $p_2 = p(\lambda_2)$ as well as the scattering phase shift $\chi = \theta_2(\lambda_1-\lambda_2)$ are parametrized by the two rapidities $\lambda_1$ and $\lambda_2$.

\subsection{Strings and magnons}
The sets of rapidities solving Bethe Eqs~\eqref{eq:betheeqprod} are self-conjugate and arrange themselves in patterns of string solutions,
\begin{equation}
  \lambda_{j,a}^{(n)}=\lambda_{j}^{(n)}+\frac{i \zeta}{2}(n+1-2a)+i \frac{\pi}{4}(1-\nu_j)+i \delta^{(n)}_{j,a}\,,
\end{equation}
where the string center $\lambda_{j}^{(n)} \in \mathbb{R}$ and $a=1,\ldots,n$ is the internal label of a rapidity within a string of length $n$ and parity $\nu_j$. In the planar regime $|\Delta| < 1$, periodicity of the trigonometric functions also allows for string centers to be located on the line $i\pi/2$, resulting in negative parity strings ($\nu_j=-1$). This type of strings will be left out of consideration for the analysis of scattering magnons, restricting to $\nu_j=1$.

At finite size, solutions are not exactly given by strings, but rather contain string deformations $\delta^{(n)}_{j,a} \in \mathbb{C}$, under the constraint that the full set of rapidities $\{\lambda\}$ remains self-conjugate. In the cases considered here the deviations are exponentially small in system size and it is therefore sufficient to take the limit of vanishing deviations. In this limit, the product of the Bethe equations of all rapidities within a string reduces to the Bethe-Gaudin-Takahashi equations,\cite{1972_Takahashi_PTP_48} which are similar to the Bethe equations but given in terms of the $n$-string centers~$\lambda_j^{(n)}$. In logarithmic form they read
\begin{multline}
  \theta_n(\lambda_j^{(n)})-\frac{1}{N} \sum_m \sum_{k=1}^{M_m} \Theta_{nm}(\lambda_j^{(n)}-\lambda_k^{(m)})=\frac{2\pi}{N}I_j^{(n)}\,,\\ 
j=1,\ldots,M_n\,,
\label{eq:BGT}
\end{multline}
where $M_n$ denotes the number of $n$-strings present, satisfying $\sum_n n M_n = M$. The logarithmic scattering kernels $\theta_n(\lambda)$ are defined in Tab.~\ref{table:kerneldef} and the scattering phase shift between two individual strings of arbitrary length is given by (for $|\Delta|<1$ we consider only strings with positive parity)
\begin{multline}
  \Theta_{nm}(\lambda) = (1-\delta_{nm})\theta_{|n-m|}(\lambda)+2\theta_{|n-m|+2}(\lambda) \\
+ \ldots + 2\theta_{n+m-2}(\lambda)+\theta_{n+m}(\lambda) \,.
\label{eq:bigthetadef}
\end{multline}
The logarithmic form of the Bethe-Gaudin-Takahashi equations allows for the introduction of string quantum numbers $I_j^{(n)}$, obeying an exclusion principle for all Bethe states, meaning that every Bethe state is characterized by a unique set of string quantum numbers. By considering the limit of sending a string center to infinity, the maximum allowed string quantum number can be derived. These limiting quantum numbers define the dimensions of sub-sectors of the Hilbert space containing a specific string content.

Moreover, in the planar case $|\Delta|<1$, the existence of strings with a specific length $n$ in the spectrum is determined by the anisotropy.\cite{1972_Takahashi_PTP_48} Therefore restrictions on the availability of $n$-string wave packets 
as well as on their momenta\cite{SutherlandBOOK}
are present in the planar case $|\Delta|<1$, depending on the value of $\Delta$.

We get all Bethe states with the desired string content by solving the Bethe-Gaudin-Takahashi equations~\eqref{eq:BGT} using an iterative algorithm for all combinations of allowed string quantum numbers. After obtaining the rapidities, the energy of a Bethe state containing strings is easily computed as
\begin{equation}
  E_{\{\lambda\}} = -\frac{J}{2}  | \phi_2(0) | \sum_n \sum_{j=1}^{M_n} \theta_n^{\prime}(\lambda_j^{(n)})\,,
\label{eq:energyfull}
\end{equation}
where $\theta_n^{\prime}$ is the derivative of $\theta_n$. The energy contribution $E^{(n)}(p)$ of a string of length $n$ to the energy $E_{\{\lambda\}}$ is
\begin{equation}\label{eq:energypartial}
  E^{(n)}(p) = J\frac{\phi_2(0)}{\phi_{2n}(0)}(\cos(p)-\epsilon_n)\,.
\end{equation}
where $\epsilon_n=\cos(n\zeta),1,\cosh(n\zeta)$ for $|\Delta|<1$, $\Delta=1$, and $\Delta>1$, respectively.
The momentum $p=p^{(n)}(\lambda)$ of an $n$-string with string center $\lambda$ is given by
\begin{equation}
  p^{(n)}(\lambda) = \pi-\theta_n(\lambda)\,.
\label{eq:stringmomentumlambda}
\end{equation}
In the following we use the convention $-\pi < p^{(n)} \leq \pi$. Note that for a single $n$-string the minimum of the energy dispersion is always at $\lambda^{(n)}=0$, i.e.~at momentum $p^{(n)}=\pi$. The total momentum of a Bethe state can be extracted from its string quantum numbers,
\begin{equation}
  P_{\{\lambda\}} = \sum_n M_n \pi - \frac{2\pi}{N} \sum_{n,j} I_j^{(n)} \mod 2\pi\,.
\end{equation}
In the thermodynamic limit and for a finite number $M$ of rapidities, each separate bound magnon represented by a string quantum number $I_j^{(n)}$ can be associated to a single particle momentum $p_j^{(n)}$
\begin{equation}
  p_j^{(n)} = p^{(n)}(\lambda_j^{(n)}) = \pi - \frac{2\pi}{N} I_j^{(n)}\,.
\label{eq:stringmomentumqno}
\end{equation}

To summarise, the rapiditites belonging to each eigenstate are obtained by iteratively solving the Bethe-Gaudin-Takahashi equations~\eqref{eq:BGT}. They can be used to evaluate the determinant expressions~\eqref{eq:matrixelements} for the normalised matrix elements of the following section.

\subsection{Magnetization expectation value}\label{sec:timeevoABA}
Time evolution of the expectation value of the local magnetization $\langle S^z_j(t) \rangle$ is performed by making use of the algebraic Bethe ansatz.\cite{KorepinBOOK} The time dependent wave function is computed using unitary time evolution in a basis of Bethe states $|\{\lambda\}\rangle$,
\begin{equation}
  |\Psi(t)\rangle = e^{-iHt}|\Psi(0)\rangle = \sum_{\{\lambda\}} e^{-i E_{\{\lambda\}} t} C_{\{\lambda\}} |\{\lambda\}\rangle \,,
\end{equation} 
where the coefficients $C_{\{\lambda\}} = \langle\{\lambda\}|\Psi(0)\rangle$ are determined by the initial state $|\Psi(0)\rangle$, which is given in Sec.~\ref{sec:wave packets}, Eq.~\eqref{eq:initial_state}, for the construction of $n$-string wave packets.

The expectation value of the local magnetization at site $j$ is given by
\begin{equation}
  \langle S^z_j(t) \rangle = \sum_{\{\lambda\},\{\mu\}} e^{-i(E_{\{\lambda\}}-E_{\{\mu\}})t} C_{\{\lambda\}} C_{\{\mu\}}^\ast \langle \{\mu\} | S^z_j | \{\lambda\} \rangle\,.
\label{eq:timeevo}
\end{equation}
As the states $|\{\lambda\}\rangle$, $|\{\mu\}\rangle$ are Bethe states determined respectively by sets of rapidities $\{\lambda_j\}_{j=1}^M$, $\{\mu_j\}_{j=1}^M$ that obey Bethe Eqs~\eqref{eq:betheeqprod} (here we have $M$ rapidities in both sets since the operator $S_j^z$ does not change the magnetization), the matrix elements are given by the normalised expressions obtained from algebraic Bethe ansatz
\begin{equation}
  \langle \{\mu\} | S^z_j | \{\lambda\} \rangle = \frac{ F^z_j(\{\mu\},\{\lambda\})}{\sqrt{\mathcal N(\{\mu\}) \mathcal N(\{\lambda\})}}\,.
\label{eq:matrixelementsnorm}
\end{equation}
Here we make use of the determinant expressions obtained in Ref.~\onlinecite{1999_Kitanine_NPB_554}, 
\begin{align}
  F^z_j (\{\mu\},\{\lambda\}) &= \frac{\varphi_{j-1}(\{\mu\})}{\varphi_{j-1}(\{\lambda\})} \prod_{k=1}^M \frac{\phi_1(\mu_k)}{\phi_1(\lambda_k)} \nonumber \\
    &\quad \cdot \frac{\det\left[H(\{\mu\},\{\lambda\})-2P(\{\mu\},\{\lambda\})\right]}{\prod\limits_{\substack{k,l=1\\ k<l}}^M \phi_0(\mu_k-\mu_l)\phi_0(\lambda_l-\lambda_k)}\, ,
\label{eq:matrixelements}
\end{align}
with $\varphi_j(\{\lambda\}) = e^{-i P_{\{\lambda\}} j}$ and $\phi_n(\lambda)$ defined in Tab.~\ref{table:kerneldef}. The entries of the matrices $H(\{\mu\},\{\lambda\})$ and $P(\{\mu\},\{\lambda\})$ are given by
\begin{align}
  H_{ab}(\{\mu\},\{\lambda\}) &= \frac{\phi_2(0)}{\phi_0(\mu_a-\lambda_b)}\Big[ \prod_{\substack{k=1\\ k\neq a}}^M \phi_2(\mu_k-\lambda_b)\nonumber \\
    &\quad -\left[\frac{\phi_{-1}(\lambda_b)}{\phi_{1}(\lambda_b)}\right]^N\prod_{\substack{k=1\\ k\neq a}}^M \phi_{-2}(\mu_k-\lambda_b)\Big]\,,
\\
  P_{ab}(\{\mu\},\{\lambda\}) &= \frac{\phi_2(0)}{\phi_{-1}(\mu_a)\phi_1(\mu_a)}\prod_{k=1}^M \phi_2(\lambda_k-\lambda_b)\,.
\end{align}
The normalization $\mathcal N(\{\lambda\})$ is computed from the Gaudin determinant,\cite{1981_Gaudin_PRD_23,1982_Korepin_CMP_86}
\begin{equation}
  \mathcal{N}(\{\lambda\}) = [\phi_2(0)]^M \prod\limits_{\substack{k,l=1\\ k \neq l}}^M \frac{\phi_2(\lambda_k-\lambda_l)}{\phi_0(\lambda_k-\lambda_l)} \det \Phi(\{\lambda\})\, ,
\label{eq:gaudinnorm}
\end{equation}
where the Gaudin matrix is given by the Jacobian of the Bethe equations,
\begin{align}
  \Phi_{ab}(\{\lambda\}) &= \delta_{ab}\Big[N\theta_1^{\prime}(\lambda_a)-\sum_{k=1}^M \theta_2^{\prime}(\lambda_a-\lambda_k)\Big]\nonumber \\
    &\quad + \theta_2^{\prime}(\lambda_a-\lambda_b)\,. 
\end{align}

The time-dependent expectation value of the local magnetization is then obtained by evaluating the double sum over matrix elements in Eq.~\eqref{eq:timeevo}. By construction, the double sum only includes eigenstates with the same particle content, which is not large for the few-magnon states we will consider. As a result, the double summation is still tractable at lattice sites $N\sim \mathcal{O}(10^2)$.  In the case of dealing with string solutions for the magnon bound states, reduced determinants for strings described in Ref.~\onlinecite{2005_Caux_JSTAT_P09003} must be used.

\section{Bound magnon wave packets}\label{sec:wave packets}
The strings described in the previous section do not correspond one to one to localized bound states of downturned spins, but rather are translationally invariant constituents of Bethe eigenstates. In order to create states of $n$ bound magnons with localized magnetization features, we construct Gaussian wave packets by summing over single $n$-string states (labeled by the string center~$\lambda^{(n)}$) with momenta distributed around $\overline p$,
\begin{equation}
  |\Psi(0)\rangle = {\cal N}_0 {\sum_{p}} e^{-ip\overline{x}-\frac{\alpha^2}{4}(p-\overline{p})^2} |\lambda^{(n)}(p)\rangle \,,
\label{eq:initial_state}
\end{equation}
where ${\cal N}_0 $ is a normalization constant. Unitary time evolution under Hamiltonian~\eqref{eq:xxzhamiltonian} implies an expression of the velocities of the wave packets by expanding the dispersion relation \eqref{eq:energypartial} around $p=\overline{p}$ to first order (see Appendix~\ref{app:sec_scattering}),
\begin{equation}\label{eq:velocity}
  v = \frac{\partial E^{(n)}(p)}{\partial p} \Big|_{p=\overline{p}} =-J\frac{\phi_2(0)}{\phi_{2n}(0)}\sin(\overline{p}) \,.
\end{equation}
The prefactor can be expanded for large anisotropy,
\begin{equation}\label{eq:velocitylargedelta}
  \frac{\phi_2(0)}{\phi_{2n}(0)} \simeq (2\Delta)^{-n+1} \,,
\end{equation}
implying that wave packets constructed from higher strings have a lower velocity in real space.

In the remainder of this section we will analyze the stability of $n$-string wave packets, along with the stability of another form of a localized multi-magnon, a consecutive-site spin block. The two constructions are closely related to each other for large $\Delta$. Subsequently, scattering processes of $n$-string wave packets are visualised by computing the time evolution of the local magnetization. Furthermore, we describe a method for direct observation of the phase accumulated by the wave function of a finite spin chain during a scattering process.

\subsection{Stability of $n$-string wave packets} \label{sec:nstringwavepackets}
The center of the wave packet $x(t)$ and its width $\Delta x(t)$ are respectively given by the expectation value and the variance of the position operator 
\begin{equation}
  \hat{x}=\sum_{j=1}^N j \left( \frac{1}{2} - S^z_j\right) \,.
\end{equation}
In the continuum limit, the sum over all possible $n$-string momenta $p=p^{(n)}$ can be approximated by an integral. For the width of the Gaussian wave packet in real space we obtain
\begin{equation}
  \Delta x(t=0):=\sqrt{ \langle\Psi(0)| \left[\hat{x}-x(0) \right]^2 |\Psi(0)\rangle} \approx \frac{\alpha}{2} \,.
\end{equation}

String solutions, being associated with bound states of magnons with exponentially decaying wave functions, will add exponential terms to the shape of the $n$-string wave packets in real space. By inserting the complex momenta of the individual constituents of the string solutions to the Bethe wave function, it can be shown that for example a 2-string state with momentum $\pi/2$ contains exponentials in the Bethe wave function reading $e^{-(x_2-x_1)/\xi}$, with an effective binding length $\xi=2/\ln(2\Delta^2)$. This average distance between the constituent particles in the string states provides a lower bound to how localized the wave packets constructed out of such string states can be.

For $\alpha<\xi(\Delta)$, the wave packet will start to lose its Gaussian shape in real space in favour of a simple exponential decay around the wave packet center. This issue can be circumvented by choosing large enough Gaussian widths, but will require much higher system sizes. The effective binding length becomes smaller at higher anisotropy, making the problem of the extra exponentially decaying shape to the Gaussian wave packet construction relevant only at $|\Delta| \lesssim 1$.

Induced by the nonlinear dispersion relation of the mag\-nons, the width of a wave packet in real space will furthermore increase in the course of time (see Eq.~\eqref{eq:appendixwidth}),
\begin{equation}
  \Delta x(t) = \sqrt{\frac{\alpha^2}{4} + \frac{\delta_n^2 t^2}{\alpha^2}} \,,
\end{equation}
obtained by expanding the dispersion relation (Eq.~\eqref{eq:energypartial}) to second order around the average momentum, yielding $\delta_n=\frac{\partial^2 E^{(n)}(p)}{\partial p^2}|_{p=\overline p}$. The broadening of $n$-string wave packets in real space is therefore described by the initial width $\alpha$ and
\begin{equation}
  \delta^2_n = J^2 \left(\frac{\phi_2(0)}{\phi_{2n}(0)}\right)^2\cos^2(\overline p)\,.
\label{eq:delta_squared}
\end{equation}
To first order, all strings of arbitrary length are stable at momenta $\overline p = \pm\pi/2$, but possess a non-trivial dependence on the anisotropy for other momenta. Moreover, the broadening of 1-string wave packets is not influenced by the anisotropy. The anisotropy dependent factor can again be approximated in the large anisotropy limit given by Eq.~\eqref{eq:velocitylargedelta}, showing that the stability of the wave packets increases with increasing anisotropy and string length.

\begin{figure}
\includegraphics{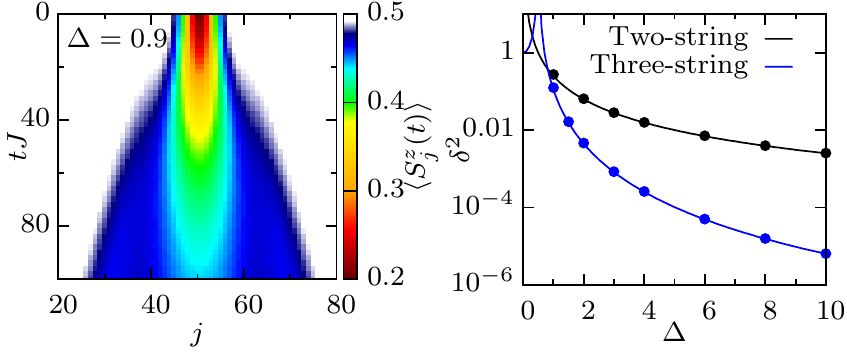}
\caption{Left: Decaying 2-string wave packet with zero group velocity (bound state of two magnons with Gaussian momentum distribution centered with $\alpha=4$ around $\overline{p}=\pi$, where the energy dispersion has its minimum), in a chain of $N=100$ sites, calculated from algebraic Bethe ansatz matrix elements. Right: prefactor $\delta^2$ of the $t^2$ dependence of the wave packet width. Theoretical curve, Eq.~\eqref{eq:delta_squared} with $\overline{p}=\pi$ and $J=1$, as a function of~$\Delta$. The data points are retrieved by fitting $\delta^2$ from the decaying wave packets from the Bethe ansatz time evolutions.}
\label{fig:lifetime}
\end{figure}

In Fig.~\ref{fig:lifetime} the magnetization profile of a diffusing 2-string wave packet with zero group velocity computed from algebraic Bethe ansatz is shown. We used as average momentum $\overline{p}=\pi$ since the energy dispersion relation has its minimum there. Furthermore, fitted parameters on the time-dependent wave packet width are compared to the theoretical values of $\delta_n^2$ for 2- and 3-strings. 

The diverging behaviour at low anisotropy of both $\delta^2_n$ and the effective binding length of strings constrain the applicability of scattering theory results for the planar regime $|\Delta| < 1$.

\subsection{Stability of spin blocks}\label{sec:stability}
\begin{figure}
 \includegraphics[width=1\columnwidth]{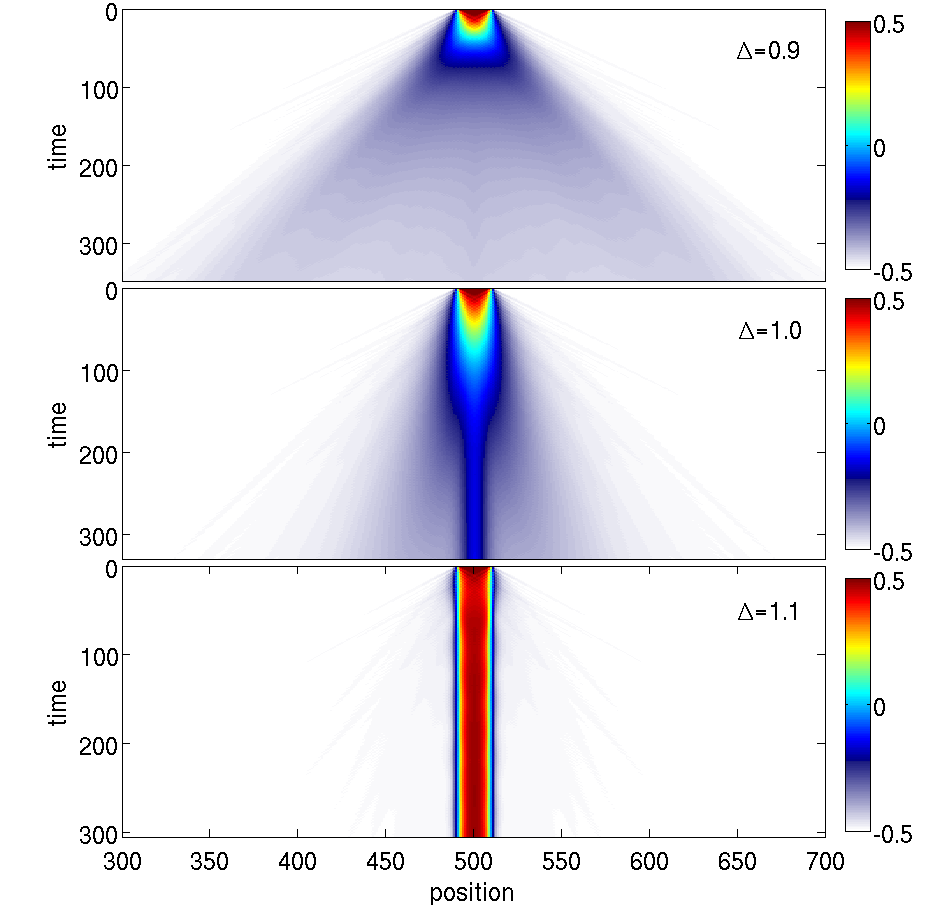}
  \caption{Decay of a block of 20 upturned spins in a chain of $N=1000$ sites for different values of anisotropy, computed using TEBD. Around the isotropic point $\Delta=1$, the behaviour of the spin block changes drastically, as the spin block becomes less tightly bound and will start to decay into smaller strings. 
}
  \label{fig:decay}
\end{figure}

A similar or even more drastic dispersing behaviour can be observed for blocks of $n$ adjacent sites. Fig.~\ref{fig:decay} shows the time evolution of a block of 20 upturned spins in a ferromagnetic chain of downturned spins for different anisotropy parameters $\Delta$. The results shown in Figs.~\ref{fig:decay} and \ref{fig:scattering} were obtained using the time evolving block decimation (TEBD) algorithm.  \cite{2012_Ganahl_PRL_108, GanahlHaqueEvertz_arxiv13,tebd}

In the axial regime $\Delta > 1$, the initial state mostly projects onto many-magnon bound states, namely $n$-strings with $n\leq 20$, which for such anisotropy values are tightly bound and thus have a large overlap with the initial spin block. A quantitative analysis of the overlap between the initial spin block and large strings is provided by Refs.~\onlinecite{2010_Mossel_JSTAT_L09001} and~\onlinecite{2010_Mossel_NJP_12} for a comparable situation involving a prepared domain wall state containing $M$ consecutive down-spins on a polarized background.  The overlaps between the $M$-spin block state and a few string configurations are considered, where the normalization saturation becomes entirely dominated by the $M$-string states with increasing anisotropy.

Since the spin blocks mostly contain large strings for large anisotropy, they display slow dispersion (see~Eq.~\eqref{eq:energypartial}), meaning that the initial spin block remains more or less intact in time over long time scales. However, as the isotropic point $\Delta = 1$ is approached, the nature of the overlaps drastically changes. Eigenstates with combinations of smaller strings start carrying a larger fraction of the total overlap with the initial state. Under time evolution, one thus sees spacetime propagation lines corresponding to shorter strings, which disperse more rapidly. By the time one has entered the planar regime, $0 < \Delta < 1$, the initial spin block decomposes into all available string lengths including the most rapidly-dispersing 1-string states, leading to a rapid dispersion of the magnetization throughout the `light cone' defined by the maximal group velocity of the 1-strings. 

A comparison of Figs~\ref{fig:lifetime} and \ref{fig:decay} 
(both at $\overline{p}=\pi$)
shows that 
spin blocks decay faster than $n$-string wave packets of the previous subsection, which has a twofold explanation. First, an $n$-string wave packet is a superposition of $n$-string Bethe states which have no decay channel into strings of smaller lengths. In contrast, in the superposition of the initial spin block state all smaller string lengths are allowed and actually present. Second, in the momentum distribution of the spin block, momenta belonging to high velocities are not Gaussian-like suppressed. Therefore, states with high velocities ($p\approx\pi/2$) are more dominant in the spin block state than in the $n$-string wave packet both with $\overline{p}=\pi$. However, the spin-block state and the $n$-string wave packets become similar at higher $\Delta$, such that the analytic predictions on scattering displacement becomes applicable to both cases.

\subsection{Scattering $n$-string wave packets}
The pre-scattering bound magnon initial states $|\Psi(0)\rangle$ are composed of two Gaussian wave packets, where the construction relies on allocation of individual momenta to distinct strings within a single Bethe state according to Eq.~\eqref{eq:stringmomentumqno}. We localize two Gaussian wave packets labeled by $j=1,2$ with average momenta~$\overline{p}_j$ around two well-separated lattice sites
\begin{equation}
  |\Psi(0)\rangle = \mathcal{N}_0 {\sum_{p_1,p_2}} c_{p_1,p_2} |\lambda^{(n)}(p_1), \lambda^{(m)}(p_2) \rangle \,,
\end{equation}
where
\begin{equation}
  c_{p_1,p_2} = e^{-i(p_1 \overline{x}_1+p_2 \overline{x}_2)-\frac{\alpha^2}{4}(p_1-\overline{p}_1)^2 -\frac{\alpha^2}{4}(p_2-\overline{p}_2)^2 }.
\label{eq:coeff}
\end{equation}
For simplicity, we label Bethe states here only by the two string centers $\lambda^{(n)}$ and $\lambda^{(m)}$ instead of the whole set $\{\lambda\}\equiv\{\lambda_j\}_{j=1}^{n+m} = \{\lambda_{1,a}^{(n)}\}_{a=1}^n \cup \{\lambda_{2,b}^{(m)}\}_{b=1}^m$. The two centers are respectively uniquely determined by the momenta $p_1$ and $p_2$ via Eq.~\eqref{eq:stringmomentumlambda}. Due to the energy dispersion \eqref{eq:energypartial} of the bound magnons, the relative velocity of the wave packets is maximized at $\overline{p}_1=-\overline{p}_2=-\pi/2$.

\begin{figure}
\centering
\includegraphics{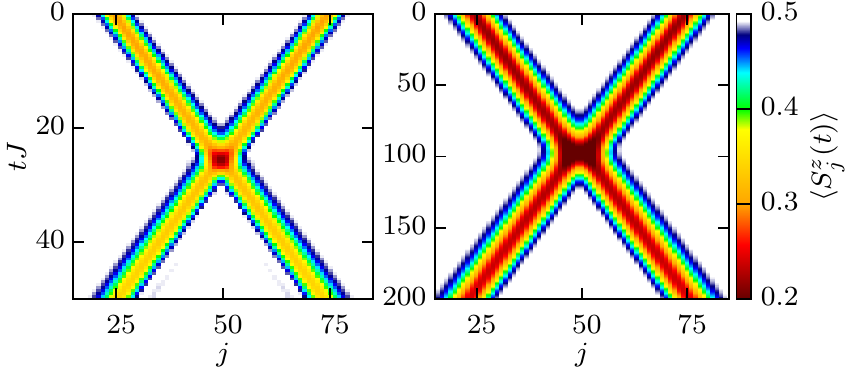}
\caption{Time evolution of $\langle S^z_j (t) \rangle$ illustrating scattering $n$-string wave packets, computed using algebraic Bethe ansatz, for $N=100$, $\Delta=2$, $\overline{p}_1=-\overline{p}_2=-\pi/2$ and $\alpha=4$. Left: scattering of 1-string wave packets (single magnons). Right: scattering of 2-string wave packets (bound magnons).}
\label{fig:scattmeas}
\label{fig:initial}
\label{fig:heatmaps}
\end{figure}

Fig.~\ref{fig:heatmaps} shows time-dependent magnetization profiles for scattering of 1- and 2-string wave packets respectively, computed from the algebraic Bethe ansatz matrix elements described in Sec.~\ref{sec:timeevoABA}. A close examination of the profiles shows distinctive features akin to soliton scattering, namely that the wave packets emerge out of a collision intact, but spatially displaced. This displacement will be quantified in the next section.

\subsection{Direct phase shift measurements}

The real-time scattering trajectories in this work are analyzed using the idea 
that the scattering between two localized wave packets is well-described by the scattering phase shift corresponding to
the average momenta of the two wave packets. This idea can also be verified through direct
observation of the phase accumulated by the wave function of a finite chain during a scattering
process. We therefore compare the evolution of an interacting chain ($\Delta\neq0$) with the evolution of a
non-interacting chain ($\Delta=0$), each containing two localized 1-string wave packets. The overlap
between the two wave functions,
\begin{equation}
  \Gamma_\Delta(t) = \langle \Psi^{[0]}(t) | \Psi^{[\Delta]}(t) \rangle \,, 
\end{equation}
gives the phase acquired due to the interaction between the magnons.  The value of $\Delta$ is here
indicated in the superscript.

In Appendix~\ref{app:phaseshiftdirect} we show how the phase of the quantity $\Gamma_\Delta$ after a
single scattering event, calculated using numerical exact diagonalization, matches the Bethe ansatz
phase shift of Eq.~\eqref{eq:BAphaseAoverA}, even for wave packets that are spatially well localized.
Such overlaps between time-evolved wave functions are currently not directly accessible by Bethe ansatz.

\section{Scattering displacement}\label{sec:displacement}

The initial state constructed from Bethe states is prepared as two wave packets of bound states of an arbitrary finite number of magnons with initial average positions and momenta $(\overline{x}_j,\,\overline{p}_j)$. The $m$- and $n$-string wave packets are constructed separately at large separation, such that their motion before and after scattering can be considered to be free. In particular, the motion of the center of each wave packet is (see Appendix~\ref{sec:app_general})
\begin{equation}
  x_j(t) = \left \{\begin{array}{c l}
    \overline{x}_j+v_j t & \textrm{before scattering,}\\
    \overline{x}_j+v_j t - \chi_j(\overline{p}_1,\overline{p}_2) & \textrm{after scattering,}
  \end{array}\right. \label{eq:displacement}
\end{equation}
given in units of lattice distance, where the velocity $v_j$ was defined in Eq.~\eqref{eq:velocity}. Note that, in the case of single magnon scattering for example, a negative average momentum $\overline{p}_j$ yields a positive velocity $v_j$ and vice versa (see Eq.~\eqref{eq:disp_rel} with $J$ positive).

The displacement $ \chi_j(\overline{p}_1,\overline{p}_2) $ can be obtained by expanding the scattering phase shift around the average momenta, see also Eq.~\eqref{eq:exp_scatt_shift}, and is therefore given as
\begin{equation}
  \chi_{j}(\overline{p}_1, \overline{p}_2) = \left. \partial_{p_j} \chi(p_1, p_2) \right|_{p_1 = \overline{p}_1, p_2=\overline{p}_2}\,,
\end{equation}
where we introduced the notation with subscript $j$ to refer to the momentum derivative $\chi_j = \frac{\partial\chi}{\partial p_j}$ of the scattering phase shift $\chi$.

Eq.~\eqref{eq:displacement} assumes that all scatterings occur without particle production, which is the case for the integrable model we are dealing with.

\subsection{Displacements from Bethe ansatz}
An analytic expression for the displacement as a function of anisotropy and incoming momenta can be extracted from the Bethe ansatz scattering phase. The phase shift of two bound magnons of arbitrary length is obtained from the scattering kernel $\Theta_{nm}$ of the Bethe-Gaudin-Takahashi Eqs~\eqref{eq:BGT}, which consists of a sum over the functions $\theta_s$, $s=|n-m|, \ldots, n+m$, defined in Tab.~\ref{table:kerneldef}, see also Eq.~\eqref{eq:bigthetadef}.

For the scattering of an $n$-string wave packet labeled by $1$ with an $m$-string wave packet labeled by $2$, the displacements on the trajectories of the $n$- and $m$-string wave packets ($j=1,2$ respectively) are given as a function of momenta as
\begin{equation}
  \chi^{(n,m)}_j(\overline p_1, \overline p_2) = \left. \frac{\partial \Theta_{nm}\left(\lambda^{(n)}(p_1)-\lambda^{(m)}(p_2)\right)}{\partial p_j}\right|_{ \substack{p_1 = \overline{p}_1 \\ p_2=\overline{p}_2 \\ \phantom{x} }}\,.
\label{eq:displdefder}
\end{equation}
The $n$- and $m$-strings with centers $\lambda^{(n)}$ and $\lambda^{(m)}$ carry momenta $\overline{p}_1$ and $\overline{p}_2$, respectively. The wave packets are located such that $\overline{x}_1 \ll \overline{x}_2 $ and $v_1 > v_2$. 

We first discuss the planar case $|\Delta| = |\cos(\zeta)| < 1$. By inverting Eq.~\eqref{eq:stringmomentumlambda}, we express the string center of an $n$-string in terms of its momentum as
\begin{equation}
  \lambda^{(n)}(p)=\text{atanh}\left(\tan\frac{n \zeta}{2}\tan\frac{\pi-p}{2}\right).
\end{equation}
Due to Eq.~\eqref{eq:bigthetadef}, expression~\eqref{eq:displdefder} for the displacement consists of a sum of the momentum derivative of the functions $\theta_s$ of which the individual terms are computed as
\begin{multline}\label{eq:kernelmomderivative}
  \frac{\partial \theta_s(\lambda^{(n)}(p_1)-\lambda^{(m)}(p_2))}{\partial p_1}=\frac{\partial \theta_s(\lambda^{(n)}-\lambda^{(m)})}{\partial \lambda^{(n)}}\frac{\partial \lambda^{(n)}(p_1)}{\partial p_1}\\
  =\frac{\sin(s \zeta)}{\cos(s \zeta)-\text{ch}(2\lambda_1^{(n)}-2\lambda_2^{(m)})} \frac{\sin(n \zeta)}{\cos(n \zeta)-\cos(p_1)}\,.
\end{multline}

For the scattering phase shift of two 1-string wave packets with $\overline{p}_1 = - \overline{p}_2 = -\pi/2$, the displacement is given as a function of anisotropy as
\begin{align}
  \chi^{(1,1)}_1\left(-\tfrac{\pi}{2}, \tfrac{\pi}{2}\right) &= \frac{\sin(2\zeta)\tan(\zeta)}{\cos(2\zeta)-\text{ch}(4\,\text{atanh}(\tan\frac{\zeta}{2}))}\notag \\
    &= \frac{1}{1+\Delta^2}-1\,.\label{eq:phase:single}
\end{align}
The latter result could have been also obtained directly by taking the derivative of Eq.~\eqref{eq:BAphaseAoverA}.

Similarly, the scattering displacement of two 2-string wave packets becomes
\begin{align}
  \chi^{(2,2)}_1\left(-\tfrac{\pi}{2}, \tfrac{\pi}{2}\right) &= \frac{2\sin(2\zeta)\tan(2\zeta)}{\cos(2\zeta)-\text{ch}(4\,\text{atanh}(\tan\zeta))}\nonumber \\
  &\quad + \frac{\sin(4\zeta)\tan(2\zeta)}{\cos(4\zeta)-\text{ch}(4\,\text{atanh}(\tan\zeta))}\notag \\
  &= \frac{16\Delta^6-4\Delta^4+5}{(4\Delta^4+1)(1+(2\Delta^2-1)^2)}-3 \,.
\label{eq:phase:bound}
\end{align}
The validity of the latter equation only extends to the region where $\Delta > 1/\sqrt{2}$, as 2-strings with momentum $p=\pm\pi/2$ do not exist for lower 
anisotropy\cite{SutherlandBOOK,2012_Ganahl_PRL_108},
which can be shown from the anisotropy dependent maximum string quantum numbers.

For the regime $\Delta > 1$, Eqs~\eqref{eq:phase:single} and~\eqref{eq:phase:bound} hold  as well, since the $\theta_s(\lambda)$ for both regimes are just rotated in the complex rapidity plane with respect to each other. Starting from $\theta_s(\lambda)$ for $\Delta>1$ with $\zeta=\text{acosh}(\Delta)$ therefore yields identical results for the scattering displacements.

The displacements for the scattering of an 1-string wave packet at a 3-string wave packet is given by
\begin{align}\label{eq:chi13_1}
  \chi^{(1,3)}_1\left(-\tfrac{\pi}{2}, \tfrac{\pi}{2}\right) &= -\frac{(4\Delta^2-1)^2(4\Delta^2-3)}{2(4\Delta^2+1)(4\Delta^4-3\Delta^2+1)} \,,\\
  \chi^{(1,3)}_2\left(-\tfrac{\pi}{2}, \tfrac{\pi}{2}\right) &= \phantom{-}\frac{(4\Delta^2-1)^3}{2(4\Delta^2+1)(4\Delta^4-3\Delta^2+1)} \,,\\
  \chi^{(1,3)}_1\left(-\tfrac{\pi}{2}, \pi \right) &= -\frac{2\Delta(2\Delta+1)^3(2\Delta^2+\Delta-1)}{(2\Delta^2+2\Delta+1)(8\Delta^4+8\Delta^3+1)} \,,\\
  \chi^{(1,3)}_2\left(-\tfrac{\pi}{2}, \pi \right) &= \phantom{-}\frac{2\Delta^2(2\Delta+1)^4}{(2\Delta^2+2\Delta+1)(8\Delta^4+8\Delta^3+1)} \,.\label{eq:chi13_4}
\end{align}

\subsection{Comparison of scattering theory and time evolution}
\begin{figure}
\includegraphics{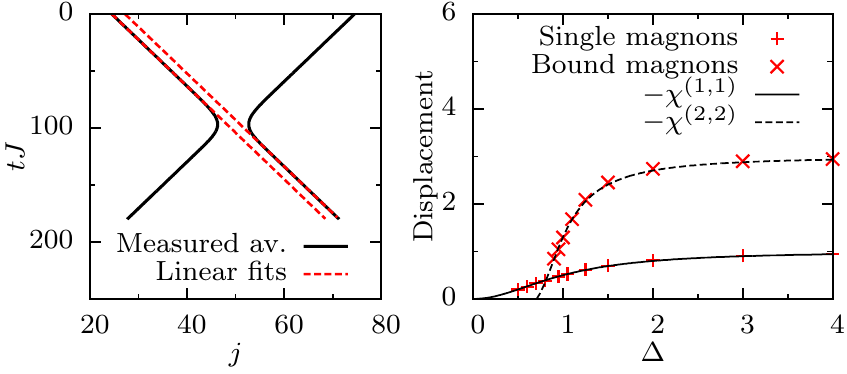}
\caption{Left: Measurement of the scattering displacement from Bethe ansatz time evolution of bound magnons at $\Delta=2$, by computing the average location of the wave packets according to Eq.~\eqref{eq:dispmeas}  and taking the horizontal difference of the linear fits. Right: Scattering displacement (in units of lattice distance) for single and bound magnons as function of anisotropy. Measured data from algebraic Bethe ansatz time evolution of magnetization (see left panel) compared to the derivative of scattering phase shifts, see Eqs~\eqref{eq:phase:single}-\eqref{eq:phase:bound}.}\label{fig:shift}
\label{fig:scatt}
\end{figure}

The displacement in the trajectories induced by scattering effects is easily deduced from the time evolution data of Fig.~\ref{fig:heatmaps}. For symmetric cases with identical particles, the average position of the magnetization of a single wave packet can be computed on one half of the system as a function of time, 
\begin{align}
  \langle j \rangle_\text{av}^\text{left}(t)&=\frac{\sum^{N/2}_{j=1} j \left(\frac{1}{2}-\langle S^z_j (t) \rangle\right)}{\sum^{N/2}_{j=1} \left(\frac{1}{2}-\langle S^z_j (t) \rangle\right)}\notag \\
    &=\frac{2}{M}\sum^{N/2}_{j=1} j \left(\frac{1}{2}-\langle S^z_j (t) \rangle\right).
\label{eq:dispmeas}
\end{align}
The result for the scattering between two $2$-string wave packets is plotted in Fig.~\ref{fig:scatt}. Note that the plotted average location of the wave packet in Fig.~\ref{fig:scatt} by means of Eq.~\eqref{eq:dispmeas} does not resemble the actual trajectories when the wave packets spatially overlap with each other, as the trajectories are not properly defined during the scattering event.
A linear fit with the same slope is applied to the propagation of the average position of the wave packet before and after scattering. The horizontal difference between the two straight lines is the displacement of the wave packet due to scattering effects. The procedure of measuring the displacements by means of Eq.~\eqref{eq:dispmeas} was performed for multiple values of anisotropy $\Delta$. Moreover, these results can be compared with Eq.~\eqref{eq:displdefder}, where the displacements for the situations of Fig.~\ref{fig:scatt} are specifically given by Eqs~\eqref{eq:phase:single} and~\eqref{eq:phase:bound} as function of anisotropy. The results are plotted in Fig.~\ref{fig:shift} as well, showing agreement between both approaches. The explicit time evolution relying on algebraic Bethe ansatz matrix elements provides confirmation of the analytical predictions for the displacements.

\begin{figure}
\includegraphics{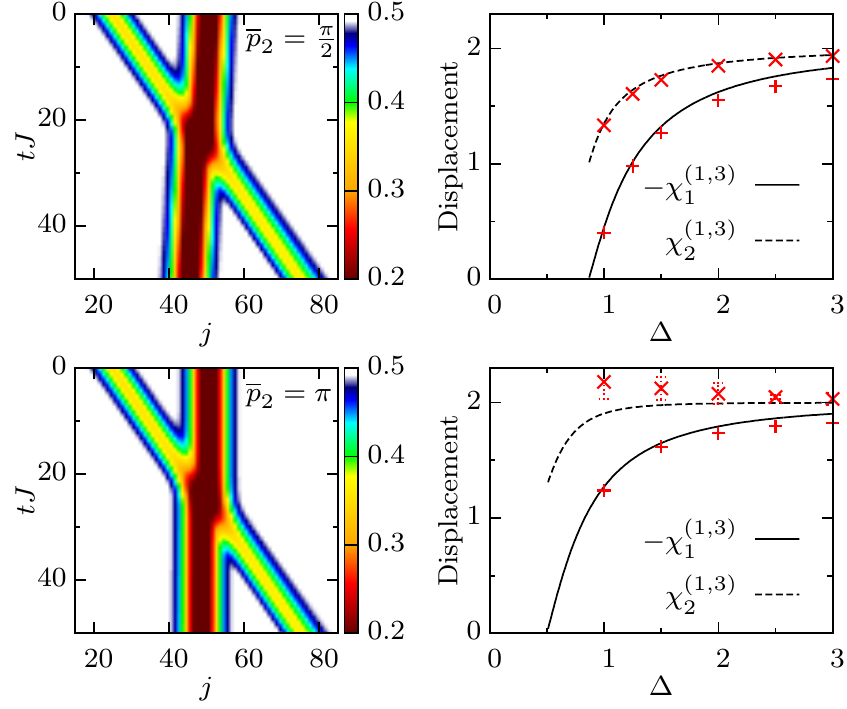}
\caption{Left: time evolution from algebraic Bethe ansatz of $\langle S^z_j(t)\rangle$ for a single magnon wave packet with momentum $\overline{p}_1=-\pi/2$ scattering against a 3-string wave packet with momentum $\overline{p}_2=\pi/2$ (top) and $\overline{p}_2=\pi$ (bottom) respectively, with $N=100$, $\Delta=2$ and $\alpha=4$. Right: Corresponding scattering displacements 
(in units of lattice distance) as function of anisotropy, see Eqs~\eqref{eq:chi13_1}-\eqref{eq:chi13_4}, where the points are measured displacements obtained from the time evolution data.}
\label{fig:multiplot_one_nstring}
\end{figure}

Besides symmetric scattering situations, colliding distinct $n$-string and $m$-string wave packets can be constructed and traced in the time-evolved magnetization profile. Fig.~\ref{fig:multiplot_one_nstring} shows the scattering between an 1- and a 3-string wave packet. Two situations are distinguished, the former with both wave packets at maximal velocity at $\overline{p}_1=-\overline{p}_2=-\pi/2$, where the larger string moves much slower because of its effective mass, see Eq.~\eqref{eq:velocitylargedelta}. The latter situation consists of an incoming 1-string wave packet scattering on a stationary wall of a 3-string wave packet at $\overline{p}_2 = \pi$. The corresponding analytic scattering displacements, Eqs~\eqref{eq:chi13_1}-\eqref{eq:chi13_4}, are shown adjacent to the time evolution plots in Fig.~\ref{fig:multiplot_one_nstring}. 
The scattering displacements are measured from the time evolution by imposing a Gaussian fit on the wave packet after scattering and comparing the average location to the time evolution of a single wave packet without scattering.
In the lower right panel ($\overline{p}_2 = \pi$), the fitting procedure of the average location of the 3-string wave packet becomes less accurate for decreasing $\Delta$.

In the planar regime where $|\Delta|<1$, we encounter substantial limitations (as described in Sec.~\ref{sec:stability}) on both the construction of the scattering wave packets, as well as on the comparison to results of scattering theory. Due to the effective binding length of the individual constituents of the string states, the tails of the magnetization profile of the wave packets start overlapping with each other significantly at low $\Delta$, invalidating important assumptions of scattering theory which include asymptotic separation of the wave packets before and after scattering. Comparison of the measured displacements to scattering theory is therefore not meaningful for higher strings in the planar regime.  Only going to much larger system sizes would resolve the aforementioned issue.

\subsection{Displacements from TEBD calculations for scattering of spin blocks compared to the Ising limit}
The scattering displacement turns out to have a particularly simple form at large anisotropy.  In particular,
it was found in Ref.\ \onlinecite{GanahlHaqueEvertz_arxiv13} that when a propagating $n$-particle
cluster is incident on a larger block, the block is displaced by $2n$ sites.  This can be explained
from the Bethe ansatz results presented in previous subsections, by taking the Ising limit $\Delta\rightarrow
\infty$ of Eq.~\eqref{eq:displdefder} for the displacement of an $n$-string wave packet
scattering at an $m$-string wave packet.

First, we obtain for all $s$
\begin{equation}
\lim_{\Delta\rightarrow \infty} \frac{\partial \theta_s\left(\lambda^{(n)}(p_1)-\lambda^{(m)}(p_2)\right)}{\partial p_1} = -1\,,
\end{equation}
which is independent of $n$, $m$, $p_1$, and $p_2$. Using Eq.~\eqref{eq:bigthetadef} for the phase shift between an $n$-string and an $m$-string eventually yields
\begin{equation}
  \lim_{\Delta\rightarrow \infty} \chi_1^{(n,m)}(\overline{p}_1,\overline{p}_2) = -2\min(n,m)+\delta_{nm}\,.
\label{eq:isingnmshift}
\end{equation}
Thus, the scattering displacement for unequal wave packets ($n\neq{m}$) is equal to twice the number
of particles in the smaller wave packet. 

The leading order term of the Ising limit in Eq.~\eqref{eq:isingnmshift} can be given for the case where $\cos(\overline{p}_1-\overline{p}_2) \neq 0$ as
\begin{align}
  \chi_{1,\text{LO}}^{(n,m)} &= -\delta_{nm} \left( (2-\delta_{n1})\frac{\cos(\overline{p}_1-\overline{p}_2)}{2} -\frac{\delta_{n1}}{2} \right) \Delta^{-2} \notag\\
    &\quad -(1-\delta_{nm})\frac{\cos(\overline{p}_1-\overline{p}_2)}{2^{|n-m|-1}}  \Delta^{-|n-m|} \notag\\
    &\quad -(2\min(n,m)-\delta_{nm})\frac{\cos(\overline{p}_1)}{2^{n-1}} \Delta^{-n} \,,\label{eq:largeDelta_correction}
\end{align}
yielding an error estimate for Eq.~\eqref{eq:isingnmshift}. A systematic expansion for all momenta becomes cumbersome due to the summation in Eq.~\eqref{eq:bigthetadef} and is left out of consideration. If $\cos(\overline{p}_1-\overline{p}_2) = 0$, the leading contribution is given by the third line of \eqref{eq:largeDelta_correction}. If further $\cos(\overline{p}_1)=0$, the leading order terms will be formed by higher powers like $\Delta^{-2n}$ for $n=m$ or $\Delta^{-2|n-m|}$ for $m\neq n$.

\begin{figure}
  \includegraphics[width=0.8\columnwidth]{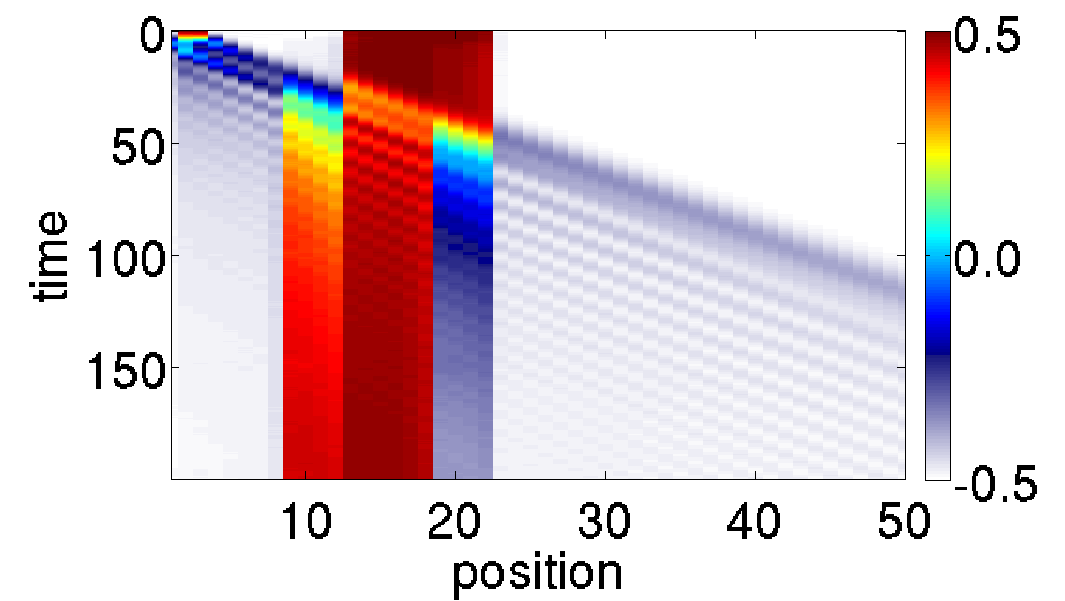}
  \caption{TEBD time evolution of scattering of a two-spin block state with a block of 10 upturned spins at $\Delta=5$. The block of 10 upturned spins is shifted by four lattice sites upon impact of the 2-string like bound magnon. This is in correspondence with Eq.~\eqref{eq:isingnmshift}, yielding a shift of four caused by a 2-string on a larger string.}
  \label{fig:scattering}
\end{figure}

Fig.~\ref{fig:scattering} shows TEBD results of the scattering of a two-spin block excitation with a
block consisting of 10 adjacent spins at $\Delta=5.0$.  The initial 2-spin block was created by
upturning two neighboring spins at lattice sites 2 and 3.  The first site (with open boundary
conditions) is energetically inaccessible for large $\Delta$; hence the 2-spin block travels to the
right and is incident on the 10-spin block.  The displacement is clearly by 4 sites, as predicted by
Eq.~\eqref{eq:isingnmshift}.  A similar displacement by 2 sites in the case of a single incident
particle was highlighted in Ref.~\onlinecite{GanahlHaqueEvertz_arxiv13}.

Although the 2-spin block and the 10-spin block are not explicity prepared as wave packets in this
case, at large $\Delta$, string states are tightly bound, and therefore these blocks may be
understood intuitively to be close to 2-string wave packets and 10-string wave packets.  The Ising
limit Eq.~\eqref{eq:isingnmshift} thus provides a satisfactory explanation to the shift observed in
these numerical experiments.  

Even more, it was observed~\cite{GanahlHaqueEvertz_arxiv13} that the behaviour of the scattering displacement of a large spin-block state close to the isotropic point at $\Delta = 1.1$ still resembles the scattering behaviour of the large $\Delta$ limit by shifting the spin-block by two lattice sites upon scattering. In order to explain this from Bethe ansatz, we take the limit of Eq.~\eqref{eq:displdefder} for the scattering of a large string ($m \gg 1$) with center $\lambda^{(m)}(p_2)$ and an 1-string with center $\lambda^{(1)}(p_1)$ at all values of $\Delta>1$,
\begin{multline}
  \frac{\partial \theta_s\left(\lambda^{(1)}(p_1) - \lambda^{(m)}(p_2)\right)}{\partial p_2} 
 \Big|_{m \gg 1} = \\[.3em]
  \left(1 + \mathcal{O}(e^{-s \zeta})\right) \left(1 + \mathcal{O}(e^{-m \zeta})\right) \,,
\end{multline}
where $s$ can only take on the values $s=m-1$ or $s=m+1$, due to Eq.~\eqref{eq:bigthetadef}. The latter equation finally gives for the displacement of the large $m$-string at $\Delta>1$,
\begin{equation}
  \chi_2^{(1,m)}(\overline{p}_1,\overline{p}_2)\big|_{m \gg 1} = 2 + \mathcal{O}(e^{-(m-1)\zeta})\, ,
\end{equation}
implying that already at $\Delta=1.1$ ($\zeta=0.4436$), the scattering displacement of a large spin block should still be close to two sites, as is in agreement with the aforementioned observation.

\section{Conclusions}\label{sec:conclusions}

In this work, we have studied the quantum analogue of soliton-like scattering phenomena in the
anisotropic spin-1/2 Heisenberg chain, by utilising the algebraic Bethe ansatz.
We considered quantum scattering of localized excitations, created from linear combinations of Bethe
states with Gaussian-distributed momenta, constructing wave packets of $n$ bound magnons.
This construction allows to study scattering phenomena of wave packets containing an arbitrary
number of bound magnons.

Exact methods based on the algebraic Bethe ansatz provide a framework to evaluate the time-dependent
expectation value of the local magnetization profile, which allows for a spatial tracking of the
localized excitations. This explicit unitary time evolution of the initial state relies on the
availability of determinant expressions for matrix elements of local spin operators.

The algebraic Bethe ansatz time evolution of colliding wave packets of bound magnons displays a spatial displacement in the trajectories of the wave packets under scattering, consistent with scattering theory results.
For different values of anisotropy, fits on the displacements of the time evolved trajectories are in agreement with analytical results on the displacement from the derivative of the Bethe ansatz scattering phase shifts, for several combinations of string lengths.

The scattering phase shift can also be measured directly as well for the scattering between two localized single-magnon wave packets, again matching phase shift expressions provided by Bethe ansatz.
Using TEBD, scattering displacements from spin-block states have been studied, showing similar
scattering features and validating the analytic predictions of the Ising limit for the scattering
displacement.
 
The experimental realizability of real time tracking of localized excitations in the Heisenberg spin
chain\cite{2013_Fukuhara_NATURE_502, 2013_Fukuhara_NaturePhysics} might provide an opportunity to
study dynamical scattering phenomena of (bound) magnons. A possible manifestation of such phenomena
might be provided by the soliton-like scattering effects analysed in this work.

The results on the scattering displacements can be extended to other Bethe ansatz solvable models.
Finally, the time evolution method relying on matrix element expressions from algebraic Bethe ansatz
can be used to construct other initial states in spin chains as well and to study their respective
relaxation phenomena.

\section*{Acknowledgements} 
We would like to thank I. Bloch, F. Essler and G. Mussardo for useful discussions. R.V., D.F., M.B., and J.-S.\,C.~acknowledge
support from the Foundation for Fundamental Research on Matter (FOM) and from the Netherlands
Organization for Scientific Research (NWO). M.G.~and H.G.E.~acknowledge support from the FWF, SFB ViCoM F4104. The authors would furthermore like to thank the MPI PKS for hospitality during the workshop ``Quantum systems out of equilibrium'' in 2013, and the CRM for their hospitality during the ``Beyond Integrability" workshop in 2015

\appendix

\section{General results from scattering theory}\label{sec:app_general}
In this Appendix, we review some general results from quantum scattering theory, emphasizing the direct connection between physically observable quantities and (derivatives of) the scattering phase shift.\cite{TaylorBOOK,1979_Zamolodchikov}

We consider an initial state  with two well-separated quasiparticles (e.g. magnons and magnon bound states) with almost well-defined positions $(\overline{x}_1,\,\overline{x}_2)$ and momenta $(\overline{p}_1,\,\overline{p}_2)$. In the asymptotic region, i.e.~when the distance between the two wave packets is much larger than the radius of the interaction, the time evolution is free. Hence, the centers of the wave packets $x_j(t)=\langle \hat{x}_j(t) \rangle$ translate rigidly. The scattering between the two particles however introduces a displacement proportional to the derivative of the  scattering phase $\chi(p_1,p_2)$. More precisely, in the asymptotic regions, the motion of the center of each wave packet is
\begin{equation}
  x_j(t) = \left \{\begin{array}{c l}
    \overline{x}_j+v_j t & \quad\textrm{before scattering,}\\
    \overline{x}_j+v_j t-  \chi_j(\overline{p}_1,\overline{p}_2) & \quad\textrm{after scattering,}
  \end{array}\right. 
\label{eq_displacement}
\end{equation}
where $\chi_j=\frac{\partial  \chi}{\partial p_j}$ is the displacement, while the group velocity $v_j$ is given by the derivative of the dispersion relation, i.e. $v_j=\frac {\partial E_j}{\partial p_j}|_{p_j=\overline{p}_j}$.

Similarly, the scattering has an effect on the width of each wave packet $\Delta x^2_j(t)=\langle
\left[ \hat{x}_j(t)-x_j(t)\right]^2 \rangle$. For Gaussian wave packets we have to first order
\begin{equation}
  \Delta x^2_1(t) = \left \{\begin{array}{c l} 
    \frac{\alpha_1^2}4 + \frac {t^2\delta^2}{\alpha_1^2} & \quad\textrm{before scatt.,}\\
    \frac{\alpha_1^2}4 + \frac{\chi_{12}^2}{\alpha_2^2} + \frac{\left(\chi_{11}-t\delta_1\right)^2}{\alpha_1^2} & \quad \textrm{after scatt.,}
  \end{array}\right.
\label{eq:appendixwidth}
\end{equation}
where $\alpha_j=\sqrt{\Delta x^2_j(t)}\Big|_{t=0}$, while $\delta_j=\frac {\partial E^2_j}{\partial p^2_j}\big|_{p_j=\overline{p}_j}$ and  $\chi_{ij}=\frac{\partial^2\chi}{\partial p_i\partial p_j}\big|_{p_i=\overline{p}_i,\, p_j=\overline{p}_j}$. An analogous formula holds for $\Delta x^2_2(t)$ as well. 

Scattering also builds up correlations between the (initially uncorrelated) Gaussian wave packets, as can be seen from the time evolution of the correlator $\Delta x_1\Delta x_2(t) = \langle \left[ \hat{x}_1(t)-x_1(t)\right]\, \left[ \hat{x}_2(t)-x_2(t)\right]\rangle$,
\begin{equation}
  \Delta x_1\Delta x_2(t) = \left \{\begin{array}{c l}
    0 & \quad\textrm{before scatt.,}\\
    \chi_{12}\left[\frac{\chi_{11}-t\delta_1}{a_1^2}+\frac{\chi_{22}-t\delta_2}{a_2^2} \right]& \quad \textrm{after scatt.} \label{eq_corr}
\end{array}\right.
\end{equation}

All the aforementioned quantities carry information about the derivatives of the scattering phase and can be in principle measured in a scattering experiment. These results hold for any one dimensional theory with stable particles. For the XXZ spin chain the stability of magnon bound states above the binding energy  threshold is preserved by the integrability of the theory.

\section{Scattering of two particles in one dimension} \label{app:sec_scattering}
In this Appendix, we review some general results for the scattering of two particles (magnons, magnon bound states, etc.),\cite{TaylorBOOK,1979_Zamolodchikov} and derive Eqs~\eqref{eq_displacement}-\eqref{eq_corr}.  For simplicity, we consider the scattering of two distinguishable particles in a continuum integrable model. The same results can be obtained for identical particles.
 
So, let us consider two particles with asymptotic momenta $p_1$ and $p_2$ and different dispersion relations $E_i(p_i)$, $i=1,2$ in an infinite volume (zero density). The statement of the coordinate Bethe ansatz, see Eq.~\eqref{eq:bethestates}, is that in the asymptotic region where the two particles are very far apart the eigenfunctions of the system are plane waves,
\begin{equation}
  \varphi_{p_1,p_2}(x_1,x_2) = \left\{ \begin{array}{c l}
    e^{i\left( p_1 x_1+p_2 \,x_2 \right)} & x_1\ll x_2 \,,\\
    S(p_1, p_2)  \, e^{i \left( p_1 x_1+ p_2 x_2 \right) } & x_1 \gg x_2\,, \label{eq_eigenfunction}
  \end{array} \right.
\end{equation}
where $S(p_1,p_2)=-e^{i \chi(p_1,p_2)}$ is the scattering matrix and $\chi(p_1,p_2)$ the scattering phase shift. At zero density, the energy is simply
\begin{equation}
  E(p_1,p_2)=E_1(p_1)+E_2(p_2)\,.
\end{equation}
Let us briefly comment on the structure of the wave function \eqref{eq_eigenfunction}. First of all, as we discussed in the main text for the XXZ model, bound states are characterized by complex conjugate rapidities, which leads to exponentially decaying terms in the Bethe wave function \eqref{eq:bethestates} with respect to the relative coordinate. This is a feature of bound states that are characterized by a center of mass coordinate. In what follows, we do not denote these exponentially decaying terms and label the bound states only with the position of the center of mass. For elementary particles, Eq.~\eqref{eq_eigenfunction} is a consequence of the conservation of energy and momentum in one dimension, 
and as such it is valid for any model with a sufficiently short-range potential. 
Instead, if one of these particles is not elementary but it is a bound state, then for a general theory the previous simple form of the wave function is not true anymore. The bound state can decay and scattering be diffractive. However, there exist models (such as the XXZ spin chain) for which the scattering is always non diffractive. Hence, the coordinate Bethe ansatz describes a complete set of asymptotic eigenfunctions, as thoughtfully discussed  in Sutherland's book.\cite{SutherlandBOOK} For such theories, bound states  cannot decay,  but are protected by integrability. 

Let us consider the scattering problem. At time $t=0$  the two particles are far apart and have (almost) well-defined positions $\overline{x}_j$ and momenta $\overline{p}_j$, $j=1,2$. Without loss of generality, we may assume that $\overline{x}_1 \ll \overline{x}_2$ and $v_1 > v_2$, where $v_j$ is the group velocity,
\begin{equation}
  v_j=\frac {\partial E_j}{\partial p_j}\bigg |_{p_j=\overline{p}_j} .
\end{equation}
When they are far apart, the two wave packets move with velocities $v_j$. If $v_1<v_2$, the evolution is always free, while for  $v_1 > v_2$ at some time the two particles become close and the interaction plays a role.

The time evolution of a two-body wave function is given by
\begin{align}
  \psi(x_1,x_2,t) &= \int \frac{dp_1}{2 \pi}\, \frac{dp_2}{2 \pi}\, \Big[C(p_1,p_2)\notag \\
    &\quad\qquad e^{-it\left(E(p_1)+E(p_2) \right)} \varphi_{p_1,p_2}(x_1,x_2) \Big] \label{eq_time}\,,
\end{align}
where
\begin{equation}\label{eq:inner}
  C(p_1,p_2)=\int dx_1\, dx_2\,  \varphi^*_{p_1,p_2}(x_1,x_2) \, \psi(x_1,x_2)\,,
\end{equation}
$\varphi_{p_1,p_2}$ is a complete set of eigenfunctions, and $\psi(x_1,x_2)$ is the initial wave function. Now, we consider an initial state which is factorized, 
\begin{equation}
  \psi(x_1,x_2)=\psi_1(x_1)\, \psi_2(x_2)
\end{equation}
where
\begin{equation}\label{eq:defPsi}
  \psi_j(x_j)=\int \frac{d p_j}{2 \pi} \, \hat{\psi}_j(p_j)\, e^{i  p_j x_j  }\,,\quad j=1,2\,.
\end{equation}
Since we are interested in computing also the spreading in time of the wave packet, we assume that the functions $\psi_j(x_j)$ are Gaussian wave packets with minimal indetermination,
\begin{equation}
  \psi_j(x_j)=\left(\frac{2}{\pi \, \alpha_j^2}\right)^{\frac 1 4} e^{i\overline{p}_j x_j}\,e^{-\frac{\left(x_j-\overline{x}_j\right)^2}{\alpha_j^2}},
\end{equation}
where $ \Delta x_j^2=\int dx_j \left(x_j-\overline{x}_j \right)^2 |\psi_j(x_j)|^2 = \alpha_j^2/4$. This assumption may be dropped if we are interested only in the displacement, Eq.~\eqref{eq_displacement}. 

Up to now, everything is exact. However, we can make the following approximations:
\begin{itemize}
  \item Since the initial state $\psi(x_1,x_2)$ is sharply peaked around $\overline{x}_1$ and $\overline{x}_2$ with $\overline{x}_1 \ll \overline{x}_2$, the only relevant contributions to \eqref{eq:inner} come from the regions around these two points. We can then use the asymptotic formula \eqref{eq_eigenfunction}  for $\varphi_{p_1,p_2} (x_1,x_2)$ with $x_1 \ll x_2$.
 Inserting Eq.~\eqref{eq_eigenfunction} into \eqref{eq:inner}, and using definition \eqref{eq:defPsi} yields
  \begin{equation}
    C(p_1,p_2) \approx \hat{\psi}_1(p_1)\hat{\psi}_2(p_2)\,.\label{eq:asympC}
  \end{equation}
  \item Since the functions $\hat{\psi}_j(p_j)$, $j=1,2$, in \eqref{eq:asympC} 
are peaked around $p_j\approx \overline{p}_j$, we can expand the dispersion relations $E_j(p_j)$ around these momenta in the integrand of Eq.~\eqref{eq_time}. 
Moreover, for $x_1 \gg x_2$ we may substitute the asymptotic expression \eqref{eq_eigenfunction} for $\varphi_{p_1,p_2}$ 
into \eqref{eq_time} and similarly expand  the scattering phase shift $\chi(p_1,p_2)$. Up to second order these expansions read
  \begin{align}
    \phantom{xxxx}E_j(p_j) &= E_j(\overline{p}_j)+v_j(p_j-\overline{p}_j) + \frac{\delta_j}{2}(p_j-\overline{p}_j)^2 \label{eq_exp_disp}
  \intertext{with $\delta_j=\frac {\partial^2 E_j}{\partial p^2_j}\big|_{p_j=\overline{p}_j}$, and}
    \phantom{xxxx}\chi(p_1,p_2) &= \chi(\overline{p}_1,\overline{p}_2) + \chi_i(\overline{p}_1,\overline{p}_2) (p_i-\overline{p}_i)\nonumber\\
      &\quad +\frac{1}{2}(p_i-\overline{p}_i) \chi_{ij}(\overline{p}_1,\overline{p}_2) (p_j-\overline{p}_j) \label{eq:exp_scatt_shift}
  \end{align}
with $\chi_i=\frac{\partial\chi}{\partial p_i}$ and $\chi_{ij}=\frac{\partial^2\chi}{\partial p_i\partial p_j}$, and we  sum over repeated indices. These expansions up to the second order are physically meaningful only if we can ignore the distortion of the wave packet, and so they are no more valid for times long enough.
  \end{itemize}
Taking advantage of these approximations, we are now in the position to derive Eqs~\eqref{eq_displacement}-\eqref{eq_corr}. Before scattering, the two wave packets propagate freely. Since they start around $\overline{x}_1$ and $\overline{x}_2$ with $\overline{x}_1 \ll \overline{x}_2$, they are centered around $x_j(t)$ with $x_1(t) \ll x_2(t)$ for small times. Thus, we can use the asymptotic formula of  $\varphi_{p_1,p_2} (x_1,x_2)$ valid for $x_1 \ll x_2$ in Eq.~\eqref{eq_eigenfunction}, and hence perform the Gaussian integrations thus obtaining the pre-scattering results, Eqs~\eqref{eq_displacement}-\eqref{eq_corr}. Similarly, long after the scattering, we have $x_1(t)\gg x_2(t)$. Thus, taking advantage of the proper asymptotic formula \eqref{eq_eigenfunction} for the eigenfunctions and the expansion \eqref{eq:exp_scatt_shift} of the scattering phase shift we also obtain the post-scattering formulas \eqref{eq_displacement}-\eqref{eq_corr}.

\section{Obtaining the phase shift from the phase of the wave function of the full chain} \label{app:phaseshiftdirect}

\begin{figure}[tb]
\centering
\includegraphics[width=0.99\columnwidth]{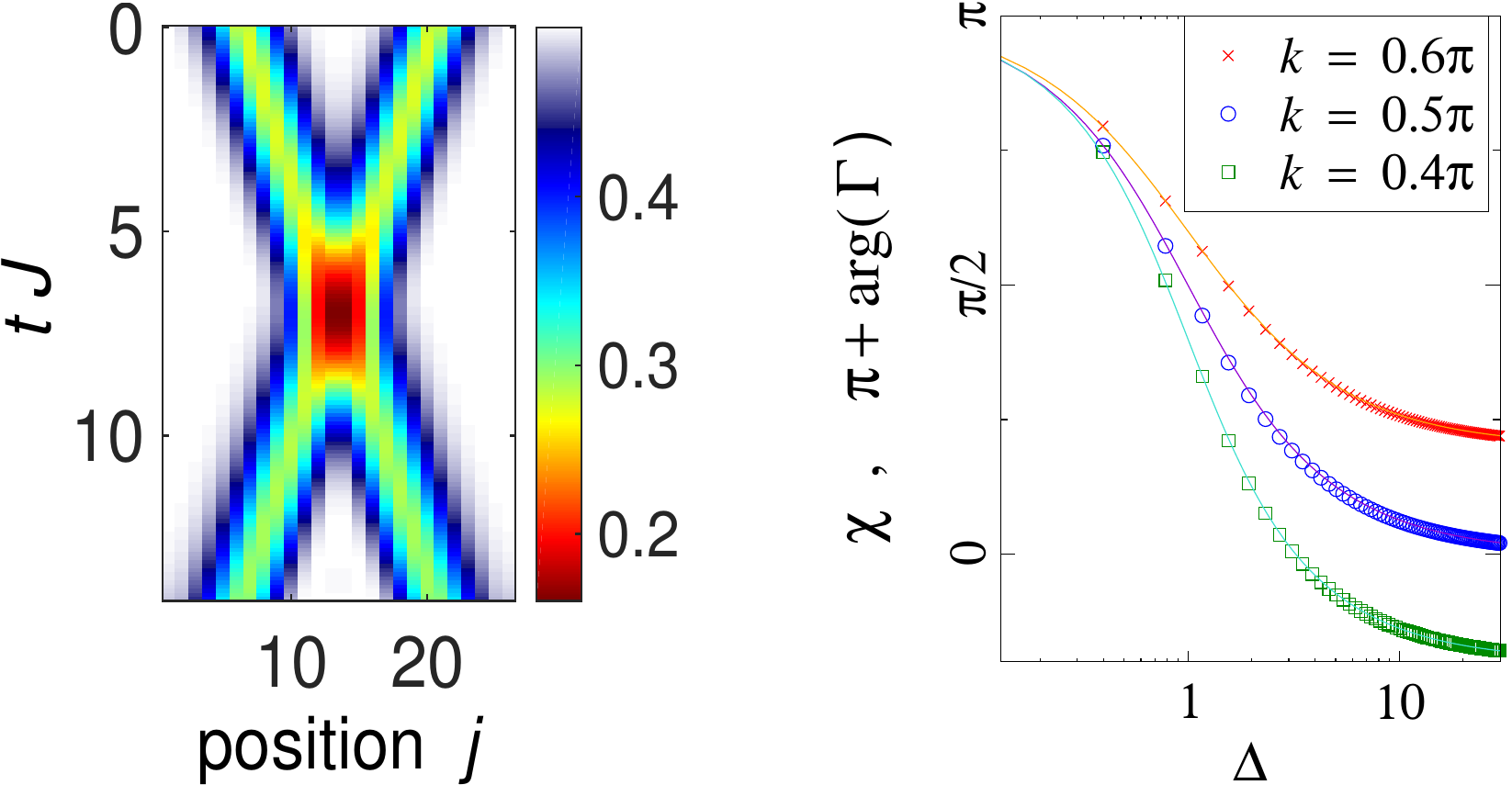}
\caption{ \label{fig:phaseshift_from_wf_phase}
Left: Time evolution of $\langle S^z_j(t)\rangle$, performed using numerical exact
  diagonalization for $L=26$ sites.  Two single-magnon wave packets are prepared with initial width
  $\alpha=5/\sqrt{2}\approx3.5355$ and opposite momenta ${\pm}k$; here $k=0.5\pi$ and
  $\Delta=2.5$.  
Right: The Bethe ansatz phase shift (lines) are compared with the phase of $\Gamma$ (symbols,
shifted by $\pi$), obtained from the wave function of the time-evolved $L=26$ chain. The comparison
is shown for three $k$ values. 
}
\end{figure}

A prominent theme of this work is that real-time scattering between two localized wave packets is
well-described by the phase shift corresponding to the average momenta of the two wave packets.  In
this Appendix we connect real-time scattering data for two single-magnon (1-string) wave packets in
a finite chain directly to the definition of the scattering phase shift, namely, that the scattering
phase shift is the phase picked up by the system wave function during the scattering process.  To
demonstrate the robustness of this idea, we show results from a stringent situation of rather small
wave packets (width of a few sites) in a rather small chain ($\approx25$ sites), far from the usual
idealized limit of infinitely extended excitations.

Here the Hamiltonian of the XXZ model is used in the following form:
\begin{equation}
H_{\Delta} ~=~ J\sum_{j=1}^{L-1}\Bigg[ S_j^xS_{j+1}^x + S_j^yS_{j+1}^y \\ ~+~
{\Delta}(S_j^z-\tfrac{1}{2})(S_{j+1}^z-\tfrac{1}{2}) \Bigg]\,.
\end{equation}
The extra terms compared to Eq.~\eqref{eq:xxzhamiltonian} are convenient for considering the phase
of the time-evolving wave function in the two-magnon sector. When evolving with the Hamiltonian
\eqref{eq:xxzhamiltonian}, there is a constant accumulation of $\Delta$-dependent phase in the time
evolution even when the magnons are spatially separated; this is avoided with the above form.  With
this form, the interaction affects the phase of the chain wave function only when the magnons meet
each other.

In the initial state, each magnon wave packet is prepared as a
Gaussian, localized approximately around $L/4$ and $3L/4$ respectively, with opposite momenta
$\pm{k}$.  The initial state is thus
\begin{multline}
  \ket{\psi(0)} ~=~ \quad {\cal N}_0 \sum_x S_x^{-} \; \exp\left[-\frac{(x-x_0)^2}{\alpha^2}\right] e^{-ikx}
  \\ \times \sum_y S_y^{-} \; \exp\left[-\frac{(y-y_0)^2}{\alpha^2}\right] e^{+iky} \quad
  \ket{\uparrow\uparrow \ldots \uparrow}  \qquad
\end{multline}
with $x_0$ and $y_0$ near the center of the left half and right half of the chain.  Here $x$ and $y$
are used as discrete site indices, and ${\cal N}_0$ is a normalization constant.  The definition of
the width $\alpha$ is chosen to be consistent with $\alpha$ in the main text.
To avoid excessive dispersion, we use $k$  near $\pi/2$.  This preparation ensures that the
two wave packets will move toward each other, collide, and continue on after the scattering, as
shown in the left panel of Fig.~\ref{fig:phaseshift_from_wf_phase}. The effect of the interaction
$J\Delta$ should be felt only when the particles are crossing each other. This interaction gives a
phase shift to the wave function during the scattering. 

The wave function at time $t$ is 
\begin{equation}
\ket{\psi^{[\Delta]}(t)} = e^{-iH_{\Delta}t} \ket{\psi^{[\Delta]}(0)}  = e^{-iH_{\Delta}t} \ket{\psi(0)} 
\end{equation}
(same initial state for every $\Delta$).   We consider the overlap 
\begin{equation}
\Gamma_{\Delta}(T) = \braket{\psi^{[0]}(T)}{\psi^{[\Delta]}(T)}
\end{equation}
at some appropriately chosen `final' time $t=T$.  Evolution with the $\Delta=0$ (non-interacting)
Hamiltonian gives a `reference' state.  
We are interested in the phase accumulated in the evolution
with the $\Delta\neq0$ Hamiltonian in comparison with the reference state, namely the phase of
$\Gamma_{\Delta}$.
The final time $T$ is chosen such that the particles have completed their scattering, but have not
reached the boundaries of the chain.  So, there are no edge effects.  For example, in the process
shown in Fig.~\ref{fig:phaseshift_from_wf_phase}(left), it would be reasonable to compare overlaps
at $T{\sim}13J^{-1}$.

The phase of $\Gamma_{\Delta}$ should approximate the scattering phase shift,
\begin{equation}
  \lim_{L\to\infty,\alpha\to\infty} \arg\left(\Gamma_{\Delta}\right)  ~=~ \chi - \pi\,. 
\end{equation}
The shift $\pi$ is in accordance with the convention used in this paper, e.g., in Eq.~\eqref{eq:BAphaseAoverA}.  
For  $k=k_2=-k_1$, the Bethe ansatz phase shift is obtained from Eq.~\eqref{eq:BAphaseAoverA} to be 
\begin{equation}
  \chi =  \pi - 2\,\atan\left(\frac{\Delta\sin(k)}{1-\Delta\cos(k)}\right)\,.
\end{equation}

Fig.~\ref{fig:phaseshift_from_wf_phase}(right) compares the phases obtained from the time-evolving
wave function of a finite chain containing two relatively narrow wave packets, with the Bethe ansatz
phase shift expressions which are strictly valid for delocalized excitations.  Even with wave
packets as narrow as $\alpha\approx3.5$, the Bethe ansatz expressions match extremely well the
phase acquired in real-time evolution.


\end{document}